\documentclass[twocolumn, switch]{article} 
\pdfoutput=1
\usepackage[colorlinks = true,
            linkcolor = purple,
            urlcolor  = blue,
            citecolor = cyan,
            anchorcolor = black]{hyperref}	
\usepackage{preprint}
\usepackage{amsmath, amsthm, amssymb, amsfonts}

\usepackage[numbers,square]{natbib}
\usepackage[utf8]{inputenc}	
\usepackage[T1]{fontenc}	
\usepackage{xcolor}		
\usepackage{booktabs} 		
\usepackage{nicefrac}		
\usepackage{microtype}		
\usepackage{lineno}		
\usepackage{float}			

\usepackage{lipsum}		

\usepackage{newfloat}
\DeclareFloatingEnvironment[name={Supplementary Figure}]{suppfigure}
\usepackage{sidecap}
\sidecaptionvpos{figure}{c}

\usepackage{titlesec}
\titlespacing\section{0pt}{12pt plus 3pt minus 3pt}{1pt plus 1pt minus 1pt}
\titlespacing\subsection{0pt}{10pt plus 3pt minus 3pt}{1pt plus 1pt minus 1pt}
\titlespacing\subsubsection{0pt}{8pt plus 3pt minus 3pt}{1pt plus 1pt minus 1pt}

\title{S-band electron spin resonance spectroscopy using a short-circuited coplanar waveguide resonator}

\usepackage{xwatermark}
\newwatermark[firstpage,color=gray!90,angle=0,scale=0.28, xpos=0in,ypos=-5in]{*correspondence: \texttt{chiranjib@iiserkol.ac.in}}

\usepackage{authblk}

\author[1, a]{Subhadip Roy}
\author[2, a]{Anuvab Nandi}
\author[1]{Pronoy Das}
\author[1]{Chiranjib Mitra}
\affil[1]{Department of Physical Sciences, Indian Institute of Science Education And Research Kolkata,India.}
\affil[2]{Department of Electronics and Telecommunication, Jadavpur University, India }

\begin{document}

\twocolumn[ 
  \begin{@twocolumnfalse} 
  
\maketitle

\begin{abstract}
In this work, we study the development of a coplanar waveguide (CPW) resonator and its use in an electron spin resonance (ESR) spectrometer. The CPW resonator is designed to operate in S-band. It has a short circuit configuration which leads to miniaturization. It is so constructed such that it has a characteristic impedance of 50 ohms. Detailed electromagnetic simulation with a particular emphasis on the excitation of the structure has been performed for this resonator owing to its uniplanar nature. The design parameters and the electromagnetic field distribution are obtained from the simulation.  The resonator is fabricated using optical lithography with a rapid prototyping technique. The characteristic response of the resonator is measured by coupling it to a Vector Network Analyzer (VNA). The ESR absorption spectrum of free radical 2,2-diphenyl-1-picrylhydrazyl (DPPH) is captured by using this resonator in reflection geometry. The microwave magnetic field distribution at the sample position is investigated.The measured g-factor value is found to be consistent with that reported in the literature. The quality factor of this resonator is found to be low and this makes it suitable for use in a Pulsed ESR spectrometer.
\end{abstract}
\vspace{0.35cm}

  \end{@twocolumnfalse} 
] 



\section{Introduction}
\label{intro}
Planar microwave resonators have versatile usage ranging from their application in dielectric measurement setup \cite{1}-\cite{2}, magnetic resonance experiments \cite{3}-\cite{4}, and as gas sensors \cite{5}.These resonators are generally based on two popular varieties of the planar transmission line, namely, microstrip line \cite{6}-\cite{mine} and coplanar waveguide (CPW) \cite{7}. A CPW is easier to fabricate when compared to a microstrip line. A CPW, being an uniplanar structure, is easier to fabricate as compared to a microstrip line. In particular, it is easier to short-circuit the signal to the ground in CPW, whereas the substrate has to be drilled and metallic vias have to be used in case of microstrip.  \cite{8}. \\
In subsequent sections, we will discuss the simulation technique used and fabrication method employed for developing a short-circuited CPW resonator for a custom-built ESR spectrometer along with its characteristics. The spectrometer is used to record the ESR spectrum of a standard free radical. 
\section{The short-circuited CPW resonator}
\label{CPW}
\subsection{Geometry}
\begin{figure}[!h]
\centering
\includegraphics[scale=0.35]{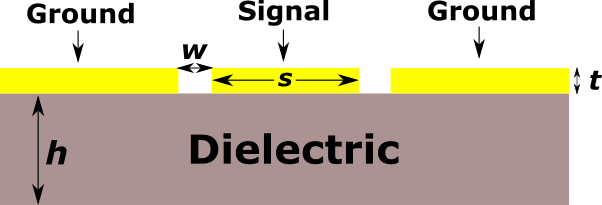}
\caption{The cross-sectional view of the CPW transmission line.}
\label{CPW1}       
\end{figure}

\begin{figure}[!h]
\centering
\includegraphics[scale=0.3]{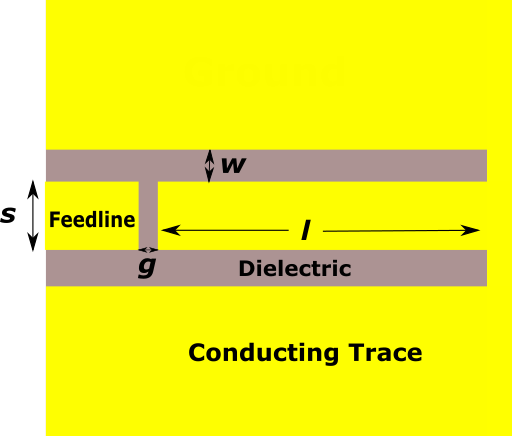}
\caption{The top view of the short-circuited CPW resonator.}
\label{res}       
\end{figure}

The planar resonator is based on an ungrounded coplanar waveguide transmission line. An ungrounded CPW transmission line consists of a dielectric of thickness \textit{\textbf{h}} and three conductive traces of thickness \textit{\textbf{t}} on top of it without any metallization on the bottom surface. The central conductive trace is the signal line of width \textit{\textbf{s}}, and the other two traces are ground lines which are at separation \textit{\textbf{w}} with respect to the signal trace. The cross-sectional view of an ungrounded CPW transmission line is shown in figure \ref{CPW1}.\\
In the resonator design, the signal and ground traces have been short-circuited. The length of the resonator \textit{\textbf{l}} is chosen so that $\text{\textit{\textbf{l}}}\approx \frac{\lambda_g}{4}$, where $\lambda_g$  is the guided wavelength corresponding to the resonance frequency $\mathit{\mathbf{f_0}}$ of the resonator. A short-circuit resonator, being quarter wavelength, is shorter than a open-circuit resonator which is half-wavelength. Therefore, a short circuit resonator leads to miniaturization. The gap \textit{\textbf{g}} separates the resonator and the feed line. The degree of coupling of the resonator to the external microwave source is controlled by \textit{\textbf{g}}. Figure \ref{res} depicts the top view of the resonator. 
\subsection{Simulation}
\begin{figure}[!h]
\centering
\includegraphics[scale=0.35]{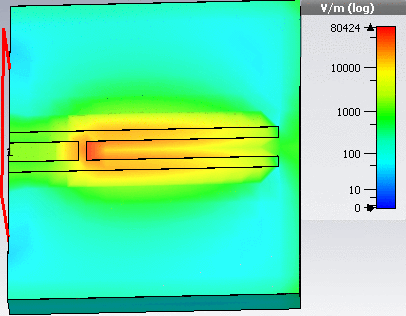}
\caption{Microwave electric field distribution of short-circuited CPW resonator.}
\label{elec}       
\end{figure}

\begin{figure}[!h]
\centering
\includegraphics[scale=0.35]{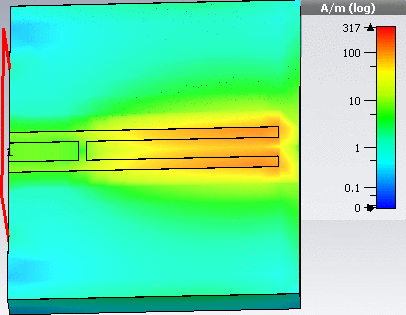}
\caption{Microwave magnetic field distribution of short-circuited CPW resonator.}
\label{mag}       
\end{figure}

The electromagnetic simulation of the resonator has been carried out with CST Microwave Studio (MWS) and ANSYS High Frequency Structure Simulator (HFSS) software. The resonator is designed to resonate at    $\mathit{\mathbf{f_0}} = 3.5$ GHz. The microwave laminate used is AD1000 (Rogers Corporation) which has a dielectric constant of 10.7 and loss tangent of 0.0023 at 10 GHz.\\
 Analytical expressions were evaluated to obtain the initial values for the design parameters \cite{9}. Parameter tuning was done in CST MWS initially to extract the design specifications . The length \textit{\textbf{l}} was tuned to achieve $\mathbf{f_0}$. The parameters \textit{\textbf{s}} and \textit{\textbf{w}} were optimized to obtain the characteristic impedance of 50 $\Omega$. The structure was excited by a waveguide port centered about the signal trace. The waveguide port was square in shape and had a side length of \textbf{k}, where \textbf{k=3(\textit{s}+2\textit{w})}. Open boundary condition is used on the side of the port, while radiating open (add space) boundary condition is used on all other sides of the resonator. The final design parameters extracted from CST MWS are tabulated in Table \ref{tab:1}. These parameters are used in fabrication. 
\begin{table}[!h]
\caption{Final design parameters }
\label{tab:1}       
\centering
\begin{tabular}{ll}
\hline\noalign{\smallskip}
Parameter & Simulated  Value  \\
\hline\noalign{\smallskip}
Length of resonator (\textit{\textbf{l}}) & 8.25 mm  \\
Coupling Gap (\textit{\textbf{g}}) & 0.33 mm \\
Signal trace Width (\textit{\textbf{s}}) & 0.9 mm \\
Gap between signal and ground traces (\textit{\textbf{w}}) & 0.5 mm \\
\textbf{Length of the feedline} & 3mm \\
\noalign{\smallskip}\hline
\end{tabular}
\end{table}
A comparative simulation of the resonator was setup in HFSS using the parameters listed in Table \ref{tab:1}. The structure was excited by a waveguide port similarly as that has been described for CST MWS.  The simulated response of the resonator obtained from both the simulation software has been compared with that of measured response in subsection \ref{fab}. While a time domain solver has been used in CST MWS, HFSS uses a frequency-domain solver, leading to a slight difference in results. The distribution of the microwave electric field and magnetic field  of the resonator for port power of 10 dBm obtained from the CST MWS simulation has been shown in figures \ref{elec} and \ref{mag}
\subsection{Fabrication and Characterization}
\label{fab}
\begin{figure}[!h]
\centering
\includegraphics[scale=0.4]{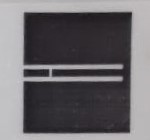}
\caption{Laser printed photomask used in the fabrication.}
\label{mask}       
\end{figure}

\begin{figure}[!h]
\centering
\includegraphics[scale=0.4]{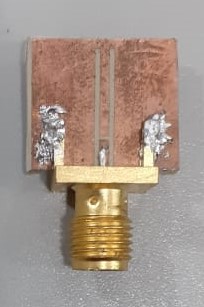}
\caption{Fabricated short-circuited CPW resonator}
\label{fab_res}       
\end{figure}

\begin{figure}[!h]
\centering
\includegraphics[scale=0.3]{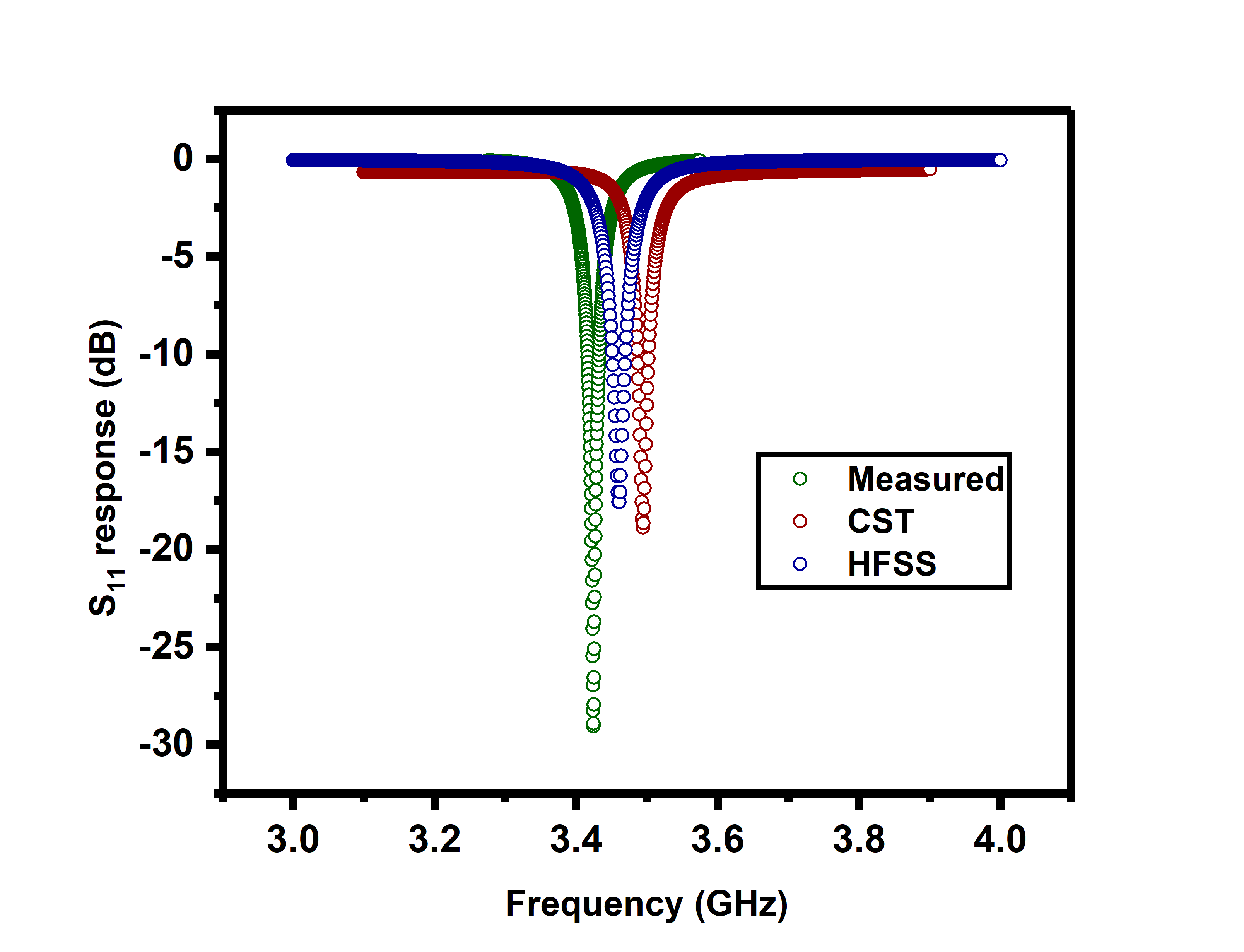}
\caption{Comparison of simulated and measured S$_{11}$ responses for 10 dBm input port power. }
\label{S11}       
\end{figure}

\begin{figure}[!h]
\centering
\includegraphics[scale=0.3]{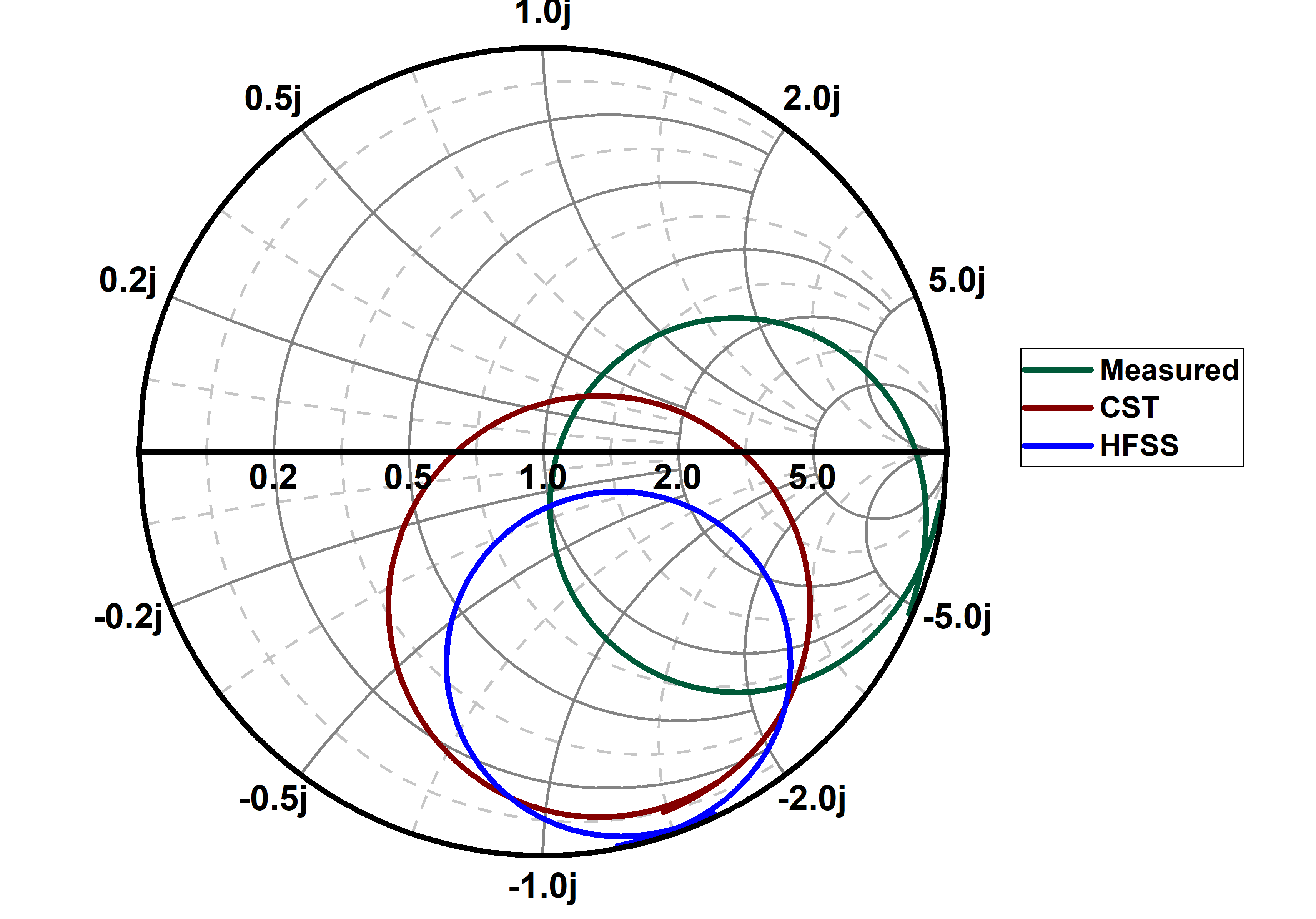}
\caption{Simulated and measured frequency sweeps of the resonator on the Smith chart.}
\label{smith}       
\end{figure}

The resonator has been fabricated on the microwave laminate using optical lithography, followed by chemical etching. The photomask is printed on tracing paper using a standard 1200 DPI Laser printer \cite{10}. A solution of de-ionized water, concentrated hydrochloric acid (35$\%$) and hydrogen peroxide (30$\%$), mixed in the ratio of 7:2:1, is used as the etchant. The overall process is cheap and fast, allowing for rapid prototyping and testing. Figures \ref{mask} and \ref{fab_res} show the used photomask and fabricated resonator, respectively. 
A vector network analyzer (VNA) (ZVA24, Rohde Schwarz) is used to measure the reflection coefficient response of the fabricated resonator in dBm. The frequency response circle is recorded on the Smith chart. The measured responses are compared with the simulation results in figures \ref{S11} and \ref{smith}.
\section{Continuous Wave Electron Spin Resonance Spectroscopy}
\begin{figure}[!h]
\centering
\includegraphics[scale=0.5]{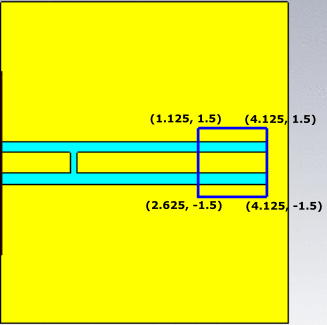}
\caption{The sample is placed inside the region indicated by the square }
\label{pos}       
\end{figure}

\begin{figure}[!h]
\centering
\includegraphics[scale=0.4]{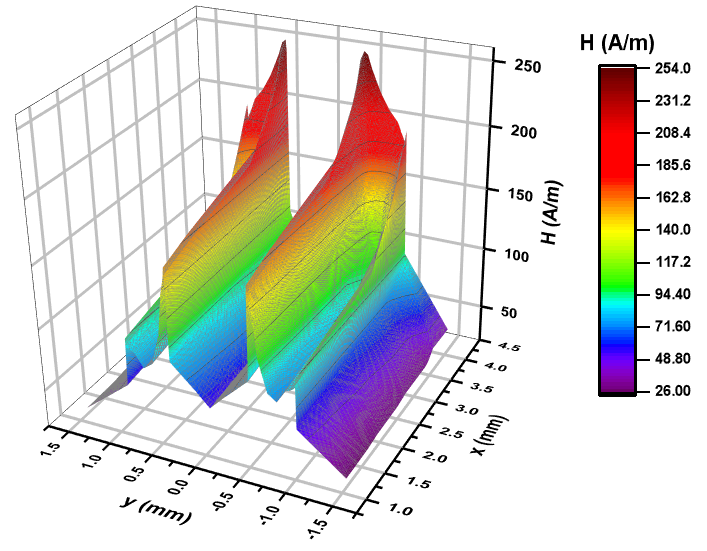}
\caption{Total magnetic field distribution in transverse plane just above the ground-signal-ground plane}
\label{mag_pos}       
\end{figure}

\begin{figure}[!h]
\centering
\includegraphics[scale=0.28]{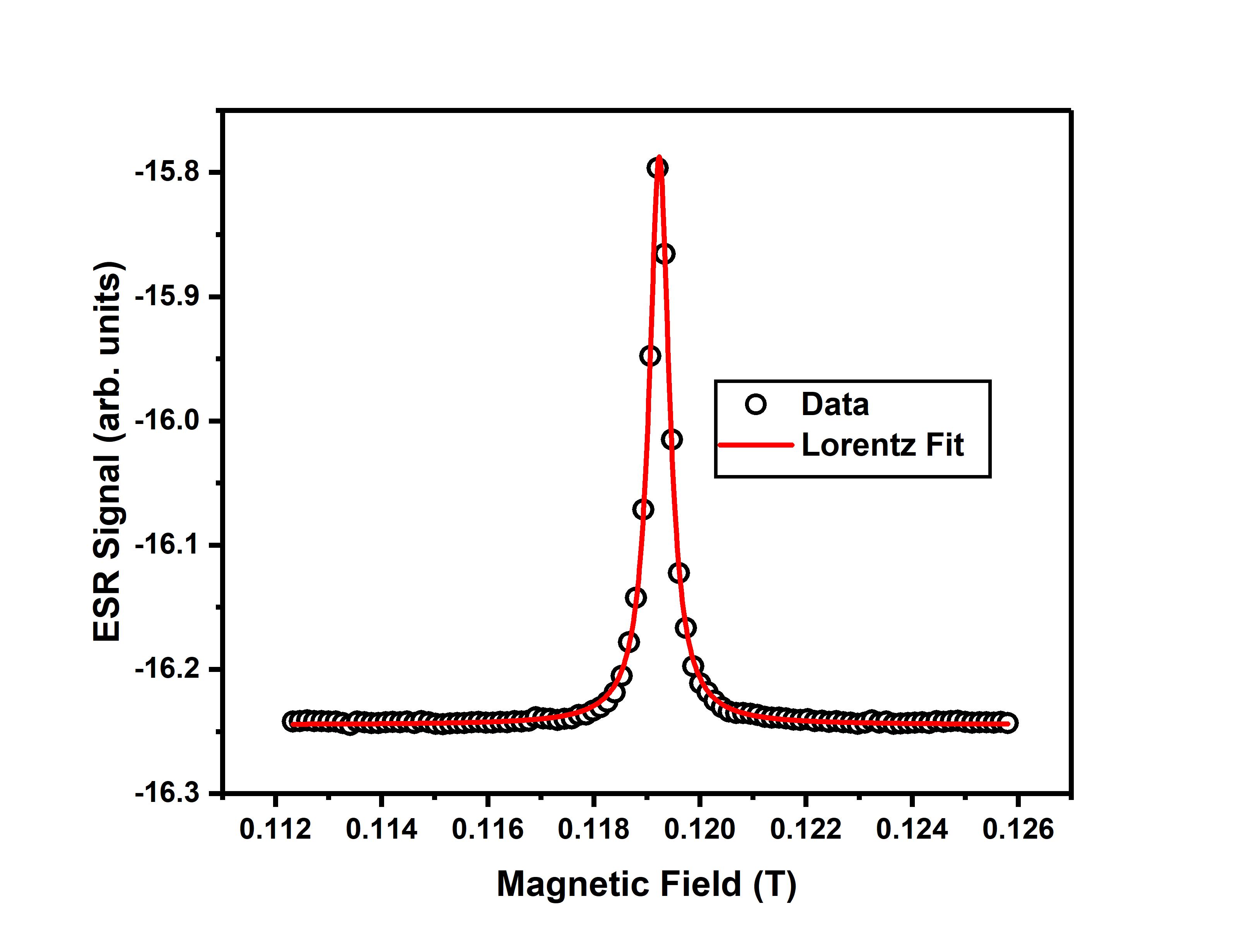}
\caption{ESR spectrum of DPPH for an input power of 10 dBm. Redline denotes the Lorentz fit.}
\label{ESR}       
\end{figure}

The fabricated resonator is used as a component of a custom-built electron spin resonance (ESR) spectrometer. The resonator is coupled to the VNA port using semi-rigid transmission line and is placed at the center of an electromagnet (GMW 3473-70). The electromagnet provides the Zeeman field. The placement of the resonator is done in such a manner so that the external magnetic field is perpendicular to the resonator’s microwave magnetic field. The magnetic field is controlled using the programmable power supply (Sorensen SGA60X83D). Calibration of the external magnetic field against the supplied current is done using a gaussmeter (DTM-151, GMW Associates). The sample 2,2-diphenyl-1-picrylhydrazyl (DPPH) is placed near the shorted end of the resonator, which is the region of the maximum magnetic field as deduced from the simulation. Figures \ref{pos} and \ref{mag_pos} indicate the sample placement region and the magnetic field distribution in that region respectively. 
The power supply of the magnet and the VNA is synchronously controlled through a custom designed digital interface. The variation in the resonant dip of the reflection coefficient of the resonator loaded with the sample is recorded against the sweep of the external Zeeman field at room temperature to obtain the ESR spectrum, as shown in figure \ref{ESR}.
\section{Results}
The measured resonance frequency of the resonator is 3.424 GHz which slightly differs from the simulations. The simulation run in CST MWS predicted a resonance frequency of 3.494 GHz and that in HFSS indicated a resonance frequency of 3.46 GHz. The S$_{11}$ response is sharper for the fabricated resonator when compared to the simulated results. The fabricated resonator has a quality factor of 70. The calculated g-factor of DPPH is 2.05, which is consistent when compared with ESR spectrometers which use planar resonators \cite{11}. Measured ESR linewidth is 4.6 Gauss. The line shape of the spectrum is Lorentzian. The signal to noise ratio (SNR) of the custom-built spectrometer is 437. 
\section{Conclusion}
The ESR spectrometer based on this resonator captures the spectra of the free radical DPPH with a decent SNR. A detailed electromagnetic simulation procedure has been described which can be adapted for any kind of CPW based resonators. The rapid prototyping technique employed here can be used to fabricate different planar transmission line based resonators \cite{mine} with reasonable accuracy. The low Q factor of the resonator makes it suitable for use in Pulsed ESR spectrometer. 
 
\section*{Authors' Contribution}
\textsuperscript{a}Both authors have contributed equally towards this work.
\section*{Acknowledgements}
Authors acknowledge Ministry of Human Resource Development (MHRD), Government of India \& Science and Engineering Research Board (SERB) (grant no. - EMR/2016/007950) for funding this work. S. R. acknowledges Council of Scientific \& Industrial Research (CSIR), India for research fellowship. The authors thank Prof. Bhaskar Gupta, Department of Electronics \& Telecommunication Engineering , Jadavpur University for providing simulation facilities. The authors are grateful to Roger Corporation, USA for free samples of the microwave laminate.


\begin{thebibliography}{}
\bibitem{1}
S. Sofin R. G. and R. C. Aiyer, Microwave and Optical Technology Letters \textbf{47}(1), 11-14 (2005). 
\bibitem{2}
R. L. Peterson and R. F. Drayton, IEEE MICROWAVE AND WIRELESS COMPONENTS LETTERS \textbf{12}(3), 90-92 (2002).
\bibitem{3} R. Narkowicz, D. Suter, R. Stonies, Journal of Magnetic Resonance \textbf{175}, 275284 (2005). 
\bibitem{4} C. Clauss, M. Dressel , and M. Scheffler, Journal of Physics: Conference Series \textbf{592} , 012146 (2015). 
\bibitem{5}  G. Bailly, A. Harrabi, J. Rossignol, M. Michel, D. Stuerga , and Pierre Pribetich , IEEE Sensors Letters \textbf{1}(4), 4500404 (2017).
\bibitem{6} L.G. Maloratsky, Microwave \& RF, March ed. 79-88(2000).
\bibitem{mine} S.Roy, S.Saha, J.Sarkar, C.Mitra, 	arXiv:2004.00457 [physics.app-ph] (2020).
\bibitem{7} C. Wen, IEEE Transactions on Microwave Theory and Techniques \textbf{17}(12), 1087-1090 (1969). 
\bibitem{8} A. Gopinath, 1979 IEEE MTT-S International Microwave Symposium Digest, Orlando, FL, USA, 1979, 109-110.
\bibitem{9} C. Nguyen, Analysis Methods for RF, Microwave, and Millimeter-Wave Planar Transmission Line Structures, Wiley-Interscience publication.
\bibitem{10}
D. Qin, Y. Xia and G.M. Whitesides, Advanced Materials \textbf{8}(11), 917-919 (1996).
\bibitem{11} W. Voesch, M.Thiemann, Daniel Bothner, M. Dressel, and M. Scheffler, Physics Procedia \textbf{75}, 503-510 (2015). 
\end{thebibliography}

\end{document}